\documentclass[pre,amsmath,amssymb,preprint,superscriptaddress]{revtex4}

\usepackage{amsmath}    
\usepackage{amsfonts}
\usepackage{amssymb}
\usepackage{graphicx}   
\usepackage{color}
\usepackage{dcolumn}
\usepackage{bm}

\begin{document}

\title{ Density jump as a function of magnetic field for collisionless shocks in pair plasmas: the perpendicular case}

\author{A. Bret}
\affiliation{ETSI Industriales, Universidad de Castilla-La Mancha, 13071 Ciudad Real, Spain}
 \affiliation{Instituto de Investigaciones Energ\'{e}ticas y Aplicaciones Industriales, Campus Universitario de Ciudad Real,  13071 Ciudad Real, Spain.}

\author{R. Narayan}
\affiliation{Harvard-Smithsonian Center for Astrophysics, Harvard University, 60 Garden St., Cambridge, MA 02138, USA}

\date{\today }

\begin{abstract}
In the absence of frequent binary collisions to isotropize the plasma, the fulfillment of the magnetohydrodynamic (MHD) Rankine-Hugoniot jump conditions by collisionless shocks is not trivial. In particular, the presence of an external magnetic field can allow for stable  anisotropies, implying some departures from the isotropic MHD jumps. The functional dependence of such anisotropies in terms of the field is yet to be determined. By hypothesizing a kinetic history of the plasma through the shock front, we recently devised a theory of the downstream anisotropy, hence of the density jump, in terms of the field strength for a parallel shock [J. Plasma Phys. (2018), vol. 84, 905840604]. Here, we extend the analysis to the case of a perpendicular shock. We still find that the field reduces the density jump, but the effect is less pronounced than in the parallel case.
\end{abstract}

\maketitle

\section{Introduction}
In a fluid shock, dissipation at the shock front is provided by binary collisions. As a result, the shock front is a few mean-free-path thick \cite{Zeldovich}. Yet, in a plasma, shockwaves can  propagate with a front far smaller than the mean-free-path. For example, the front of the earth bow shock is about 100 km thick, whereas the proton mean-free-path at the same location is of the order of the Sun-Earth distance \cite{PRLBow1,PRLBow2}. Here, dissipation is provided by collective plasma effects \cite{Sagdeev66}. Because spatial distances involved in the physics of such shockwaves are smaller than the mean-free-path, these shocks have been called ``collisionless shocks''.

Since the mechanism sustaining these shocks is different from fluid shocks, one could ask to which extent they fulfill the Rankine-Hugoniot relations (RH) for fluid or MHD shocks. These relations eventually rely on two assumptions : 1) All the matter upstream goes downstream and 2) binary collisions isotropize the distribution function on short time scales. If both assumptions are fulfilled, conservation equations can be written between the upstream and the downstream, and an isotropic equation of state can be used, ensuing RH. In the case of collisionless shocks, 1) is no longer obvious as some upstream particles may get reflected at the front while others may travel from the downstream to the upstream. Studies conducted so far in this respect found a few percent deviation from RH for the density jump \cite{Stockem2012}, and up to a few tens percent for the downstream temperature \cite{Caprioli2014}.

Regarding the  assumption 2), namely that the distribution function is isotropized on short time scales, it has been known for long that the presence of an external magnetic field can jeopardize it \cite{CGL1956}. Particle-In-Cell (PIC) simulations recently conducted \cite{BretJPP2017} found a significant reduction of the density jump for the case of a flow-aligned magnetic field, as the field tends to guide the particles downstream \cite{BretJPP2016}, preventing them to isotropize. Indeed, in the presence of an external field, the Vlasov equation does not impose an isotropic distribution function. It simply limits the range of stable anisotropies instead, through the mirror or the firehose instabilities for example \cite{Gary1993}. Notably, the kinetic theory sustaining these results has been beautifully verified in the solar wind \cite{MarucaPRL2011,SchlickeiserPRL2011}.

Several authors already studied how an anisotropic pressure in the downstream affects the RH jump conditions \cite{Karimabadi95,Erkaev2000,Vogl2001,Gerbig2011}. However, the downstream anisotropy is considered a free parameter in these works. Our objective is precisely to compute it in terms of the field strength.

In order to devise a theory of the density jump in terms of the field, we recently implemented a model for the case of a magnetic field parallel to the flow \cite{BretJPP2018}. We considered a pair plasma, for simplicity in such an exploratory work. Electron/ion plasma could add a layer of complexity to the problem, be it because both species might be heated differently at the front \cite{Guo2017,Guo2018,2019NatAstro}. The firehose and the mirror instabilities in pair plasmas have been found similar to the ones in electron/ion plasmas \cite{Gary2009}, allowing us to use the same stability criteria.

Since the Vlasov equation alone does not impose a unique value of the downstream anisotropy, we made some hypothesis on the thermal history of the plasma through the front. In our previous work on parallel shocks, we assumed the following \cite{BretJPP2018},
\begin{itemize}
  \item The upstream is isotropic.
  \item As it passes through the front, the plasma conserves its temperature perpendicular to the motion. This assumption stems from the double adiabatic theory of Chew-Goldberger-Law \cite{CGL1956}. The additional entropy generated at the front goes into the direction parallel to the motion since this is the direction of the compression. We labelled ``Stage 1'' this first stage of the plasma history.
  \item If ``Stage 1'' is stable, then this is the final state of the downstream.
  \item If ``Stage 1'' is unstable, then the plasma migrates to the stability threshold (firehose). This is ``Stage 2'', which is therefore reached only if ``Stage 1'' is  unstable.
  \item In each case, the conservation equations entirely determine the downstream parameters, density jump included.
\end{itemize}

The goal of this work is to extend the analysis to the perpendicular case.
The model was non-relativistic. We found that for an adiabatic index $\gamma=5/3$, the density jump in the strong field limit reaches 2 whereas the corresponding MHD value is 4. Notably, for a flow-aligned field, the fluid disconnects from the field in MHD so that the shock properties are independent of the field \cite{Lichnerowicz1976}. Any change in the system when varying the field can therefore be related to a deviation from the MHD behavior.

  \begin{figure}
  \begin{center}
   \includegraphics[width=.45\textwidth]{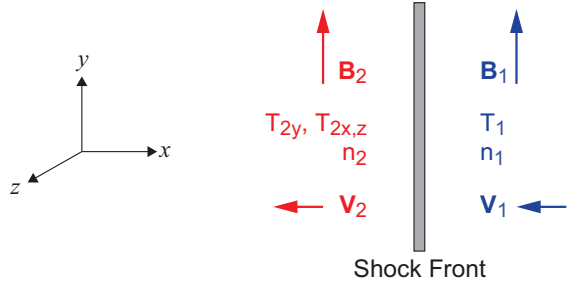}
  \end{center}
  \caption{System considered. The upstream is isotropic. The downstream is anisotropic with $T_{2y} \neq T_{2x,z}$. Due to the orientation of the field, the Vlasov equation imposes $T_{2x}=T_{2z}$.}\label{fig:setup}
 \end{figure}

The system considered in the present work is pictured in Figure \ref{fig:setup}. The plasma comes from the right and goes leftward. Upstream quantities all bear the subscript ``1'', and downstream quantities the subscript ``2''.  In order to avoid confusion, we shall not qualify pressures or temperatures with the adjectives ``parallel'' or ``perpendicular'' but will refer to the axis $x,y,z$ instead.

``Stage 1'' is still  defined by a downstream plasma having the same perpendicular (to the motion) temperature than the upstream, that is, $T_{2y}=T_1$. We still assume that the excess entropy generated at the front crossing goes into the $x,z$ directions. Notably, the Vlasov equation imposes a gyrotropic distribution around the field so that $T_{2x} = T_{2z}$ (see for example Ref. \cite{LandauKinetic}, \S 53). As will be showed in the sequel, ``Stage 1'' can be mirror unstable. In case it is, the downstream plasma therefore migrates toward the mirror instability threshold, that is, ``Stage 2''. Whether we deal with Stage 1 or Stage 2, the conservation equations fully determine the state of the downstream, hence, the density jump.

The article is structured as follows. We start reminding the results of the isotropic MHD theory in Section \ref{MHD}. The properties and mirror stability of Stage 1 are assessed in Section \ref{sec:stage1}. Section \ref{sec:stage2} then characterizes the state of the downstream in case it has to move to Stage 2, on the mirror instability threshold. A global picture of the density jump in terms of the field is finally presented in Section \ref{sec:together}, before we reach our conclusions.

\section{Formalism and MHD results}\label{MHD}
Contrary to the parallel case, the field is not conserved across the shock in MHD and enters the non-relativistic conservation equations \cite{Kulsrud2005},
\begin{eqnarray}
  n_1 V_1                                  &=&  n_2 V_2,                                  \label{eq:conser1} \\
  n_1 V_1^2 + P_1  + \frac{B_1^2}{8\pi}    &=&  n_2 V_2^2 + P_{2x} + \frac{B_2^2}{8\pi} , \label{eq:conser2} \\
  V_1 B_1                                  &=&  V_2 B_2 ,                                   \label{eq:conser3} \\
  \frac{V_1^2}{2} + \frac{P_1}{n_1} + U_1 + \frac{B_1^2}{4\pi n_1} &=&  \frac{V_2^2}{2} + U_2 + \frac{P_{2x}}{n_2} + \frac{B_2^2}{4\pi n_2}, \label{eq:conser4}
\end{eqnarray}
where $U$ is the internal energy of the fluid. As in the parallel case, the downstream pressure entering the equations is the one along the $x$ axis, which is the direction of the fluid motion (see Ref. \cite{FeynmanVol2}, \S 40-3). Here, the direction of motion and the field are perpendicular.

We now introduce the dimensionless parameters,
\begin{equation}\label{eq:dimless}
  r  = \frac{n_2}{n_1}, ~~~  A_2 = \frac{T_{2x,z}}{T_{2y}}, ~~~\chi_1^2 = \frac{V_1^2}{P_1/n_1}, ~~~ M_{A1}^2 = \frac{n_1 V_1^2}{B_1^2/4\pi},
\end{equation}
where the downstream anisotropy reads $A_2 = T_{2x}/T_{2y} = T_{2z}/T_{2y}$. The $\chi_1$ parameter looks like a Mach number, but since we force the degrees of freedom of the plasma, it is preferable to deter such an interpretation to the end of the analysis.

In order to make the junction with PIC simulations (see \cite{Marcowith2016} and references therein), and with our previous treatment of the parallel case, we also introduce the dimensionless parameter,
\begin{equation}\label{eq:sigma}
  \sigma =  \frac{B_1^2/4\pi}{n_1 V_1^2} = \frac{1}{M_{A1}^2}.
\end{equation}
We now remind the MHD results for a perpendicular shock.

\begin{figure}
\begin{center}
 \includegraphics[width=.45\textwidth]{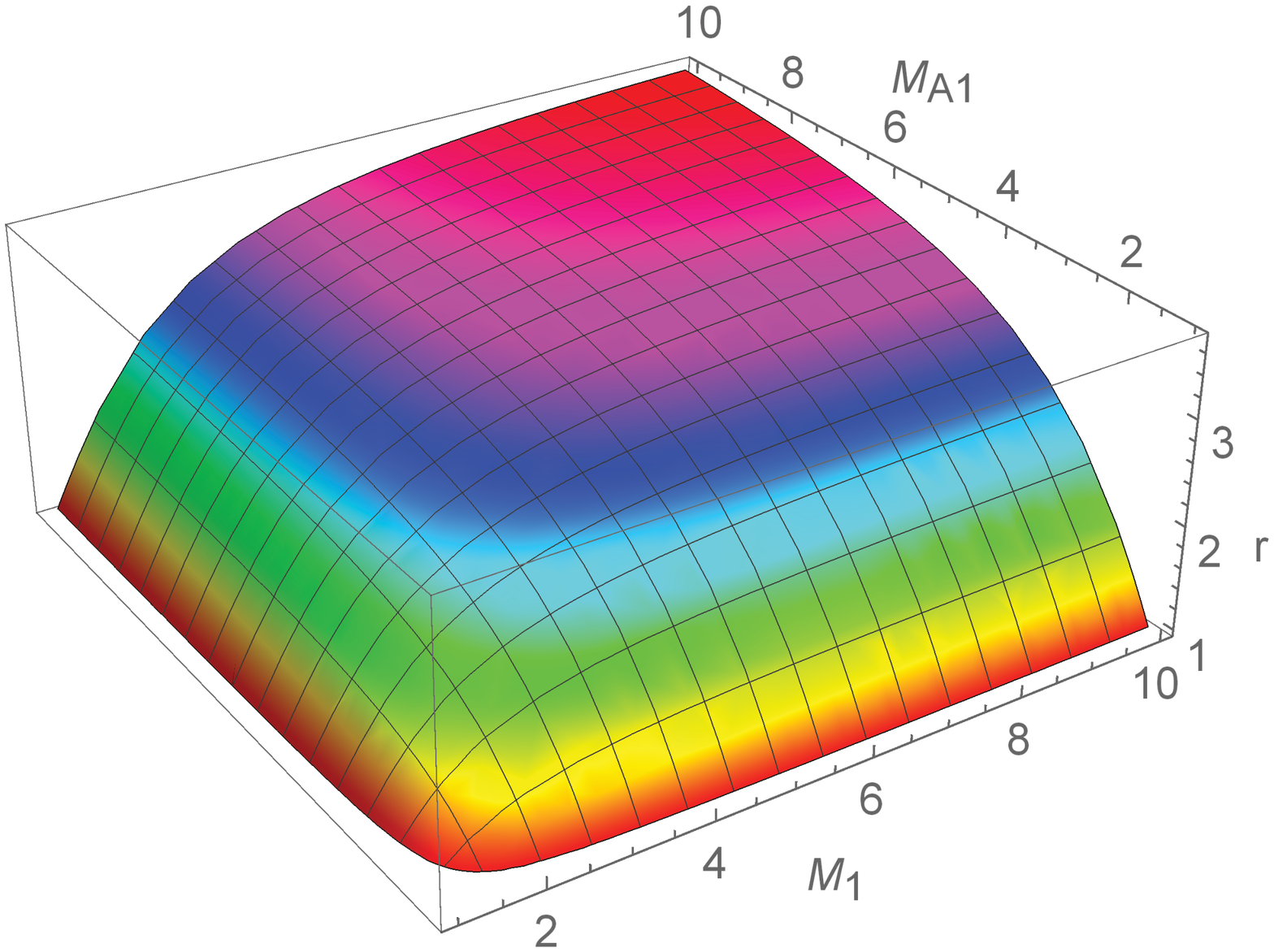} \includegraphics[width=.45\textwidth]{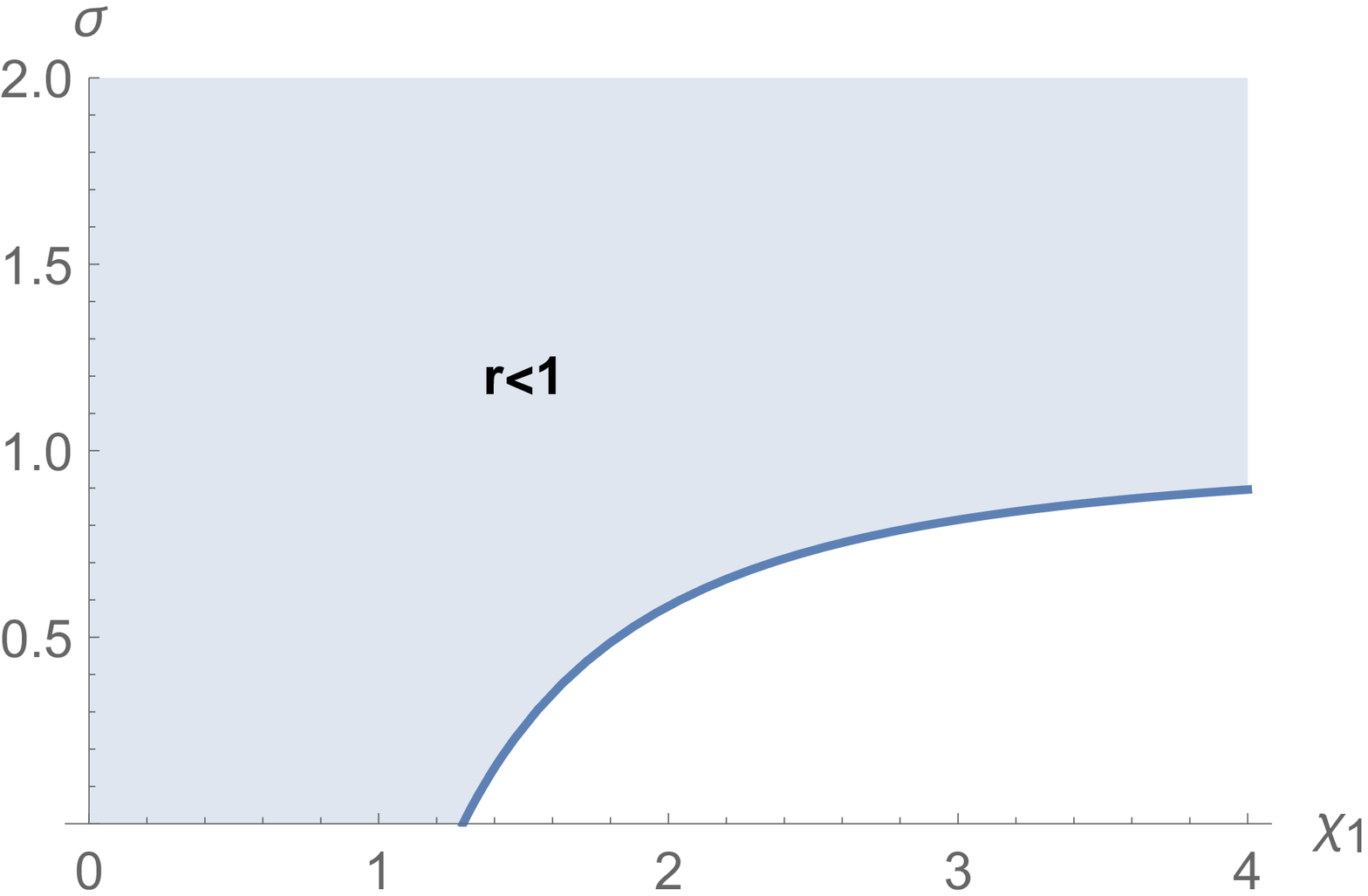}
\end{center}
\caption{\emph{Left}: Plot of the MHD density jump (Eq. \ref{eq:jumpMHD}) in terms of the upstream Mach number $M_1$ and Alfv\'{e}n Mach number $M_{A1}$ for $\gamma=5/3$, and over the domain defined by (\ref{eq:frontierMHD}). \emph{Right}: Domain of the $(\chi_1,\sigma)$ phase space yielding shock solutions according to Eq. (\ref{eq:frontierMHD}) and for $\gamma=5/3$.}\label{fig:jumpMHD}
\end{figure}

\subsection{MHD results}
 We here set $U_i=P_i/n_i(\gamma-1)$. From Eq. (\ref{eq:conser2}), one can express $P_{2x}$ in terms of $n_2$ and the upstream quantities (also using Eqs. \ref{eq:conser1} and \ref{eq:conser3}). We then do the same with Eq. (\ref{eq:conser4}), and equate the 2 expressions of $P_{2x}$. The resulting equation for $n_2$ is a 3rd degree polynomial in $r$. It can be factored by $(r-1)$ since the conservation equations obviously admit plasma continuity as a solution. The remaining 2nd order polynomial has one negative root. The positive one is \cite{fitzpatrick2014plasma},
\begin{eqnarray}\label{eq:jumpMHD}
  r &=& \frac{\gamma  M_1^2+M_{A1}^2 \left(  2+(\gamma -1) M_1^2 \right) -\sqrt{\Delta} }{2 (\gamma -2) M_1^2}, ~~\mathrm{with} \nonumber \\
  \Delta &=& 4  (\gamma-\gamma^2+2 ) M_1^4 M_{A1}^2 \nonumber \\
           && +\left[\gamma  M_1^2+M_{A1}^1 \left(2+(\gamma -1) M_1^2\right)\right]^2,
\end{eqnarray}
where we defined the upstream Mach number $M_1^2=n_1V_1^2/\gamma P_1 \equiv V_1^2/C_{s1}^2$ ($C_{s1}$ is the upstream speed of sound).

The density jump is larger than unity for,
\begin{equation}\label{eq:frontierMHD}
M_1^2 > \frac{M_{A1}^2}{\sqrt{M_{A1}^2-1}}   \Rightarrow    V_1^2 > C_{s1}^2 + V_{A1}^2,
\end{equation}
with $V_{A1}^2 = V_1^2/B_1^2/4\pi n_1$. Figure  \ref{fig:jumpMHD}-\emph{left} plots Eq. (\ref{eq:jumpMHD}) over the domain defined by Eq. (\ref{eq:frontierMHD}).

In terms of the parameters $\chi_1,\sigma$ defined by Eqs. (\ref{eq:dimless},\ref{eq:sigma}), we find,
\begin{equation}\label{eq:NoShockMHD}
r < 1 ~~~ \mathrm{for} ~~~ \sigma > \frac{\chi_1^2-\gamma }{\chi_1^2},
\end{equation}
so that the shock solutions are limited by the strength of the field. In order words, too strong a field switches off the MHD shock. Figure \ref{fig:jumpMHD}-\emph{right} pictures the portion of the $(\chi_1,\sigma)$ phase space yielding shock solutions for $\gamma=5/3$. The requirement $r>1$ imposes $\chi_1^2>\gamma$, that is, $M_1>1$.

\section{Stage 1: Downstream with $T_{2y}=T_{1y}=T_1$}\label{sec:stage1}
Since the upstream is considered isotropic, we simply set $U_1=P_1/n_1(\gamma-1)$ in Eq. (\ref{eq:conser4}). In addition, we consider $\gamma=5/3$ in the sequel.

As specified earlier, our ansatz is that $T_y$ is conserved when crossing the front while the entropy increase goes into the $T_{x,z}$ directions.
In order to express $U_2$ accounting for this ansatz, we start from,
\begin{equation}\label{eq:U}
  U_2 = \frac{1}{2n_2}(P_{2y}+P_{2x}+P_{2z}) = \frac{1}{2}(k_BT_{2y}+2k_BT_{2x}).
\end{equation}
Using $T_{2y}=T_1$, we get
\begin{equation}
  U_2 = \frac{1}{2}(k_BT_1+2k_BT_{2x}) = \frac{1}{2}\left(\frac{P_1}{n_1}+2\frac{P_{2x}}{n_2}\right).
\end{equation}
We now insert this expression into Eq. (\ref{eq:conser4}), and apply the resolution method described for the simple MHD case. We find only 2 solutions for the jump. One is $r=1$ and the second is,
\begin{eqnarray}\label{eq:rCGL}
  r&=&\frac{3 M_{A1}^2 \chi_1^2}{M_{A1}^2 \left(\chi_1^2+4\right)+2 \chi_1^2}, \nonumber \\
   &=& \frac{3 \chi_1^2}{(2 \sigma +1) \chi_1^2+4}.
\end{eqnarray}
At low $B_1$ (high $M_{A1}$), the corresponding strong shock has $r=3$, which corresponds to a strong 2D shock. Then increasing $B_1$ lowers $r$. This jump is larger than unity for
\begin{equation}\label{eq:fronteer}
\chi_1^2 > \frac{2 M_{A1}^2}{M_{A1}^2-1} \Rightarrow V_1^2 > V_{A1}^2 + 2 C_s^2.
\end{equation}

A notable consequence of Eq. (\ref{eq:rCGL}) for the density jump is that,
\begin{equation}\label{eq:NoShock}
r < 1 ~~~  \mathrm{for} ~~~  \sigma > \frac{\chi_1^2-2}{\chi_1^2},
\end{equation}
clearly reminiscent of Eq. (\ref{eq:NoShockMHD}), the corresponding relation for the MHD case. Our ansatz for Stage 1 eventually leaves the downstream plasma with 2 degrees of freedom, hence an effective adiabatic index of 2.  Eq. (\ref{eq:NoShock}) is therefore coherent with Eq. (\ref{eq:NoShockMHD}), and the strong shock limit of Stage 1, namely $r=3$, is also coherent with the effective $\gamma$.

We need now compute the Stage 1 anisotropy in order to assess its stability. Some algebra shows that,
\begin{equation}
A_2=\frac{T_{2x,z}}{T_{2y}} = \frac{P_{2x}/n_2}{P_1/n_1} = \frac{1}{r}\frac{P_{2x}}{P_1},
\end{equation}
where $P_{2x}$ is computed from Eq. (\ref{eq:conser2}). The result is,

\begin{equation}\label{eq:A2CGL}
A_2=\frac{1}{r}-\frac{(r-1) \chi_1^2  (r^2+r-2 M_{A1}^2 )}{2 M_{A1}^2 r^2}.
\end{equation}

\begin{figure}
\begin{center}
 \includegraphics[width=.45\textwidth]{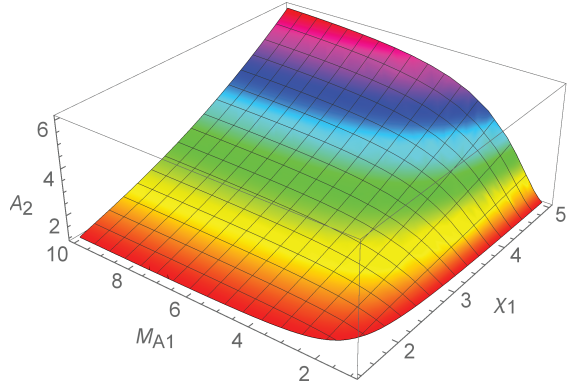}
\end{center}
\caption{Plot of the anisotropy $A_2$ given by Eq. (\ref{eq:A2CGL}) over the range defined by (\ref{eq:fronteer}).  $\gamma=5/3$.}\label{fig:A2CGL}
\end{figure}

Clearly,  $r=1$ gives $A_2=1$, so that $A_2=1$ on the frontier defined by (\ref{eq:fronteer}). Figure \ref{fig:A2CGL} plots the anisotropy $A_2$ given by Eq. (\ref{eq:A2CGL})   over the domain defined by (\ref{eq:fronteer}). We have $A_2>1$. As a consequence, Stage 1 could be mirror unstable.

\subsection{Mirror stability of Stage 1}\label{sec:mirror1}
The threshold for the mirror instability is defined by \cite{Gary1993},
\begin{equation}\label{eq:BetaPara2}
\frac{T_{2\perp}}{T_{2\parallel}} = A_2 = \frac{T_{2xz}}{T_{2y}} = 1+\frac{1}{\beta_{2\parallel}},
\end{equation}
where the subscripts $\parallel$ and $\perp$ refer to parallel and perpendicular to the field. The parameter $\beta_{2\parallel}$ can be expressed from,
\begin{equation}\label{eq:BetaPara}
\beta_{2\parallel} = \frac{n_2T_{2\parallel}}{B_2^2/8\pi}= \frac{n_2T_{2y}}{B_2^2/8\pi} = \frac{2}{r\sigma \chi_1^2}.
\end{equation}
Eqs. (\ref{eq:BetaPara2},\ref{eq:BetaPara}) yield a stability condition for Stage 1 defined by the following 3rd degree equation in $\sigma$,
\begin{equation}
A_2 =  1+\frac{r\sigma \chi_1^2}{2} ~~ \Leftrightarrow  ~~  \sum_{k=0}^3a_k\sigma^k =0,
\end{equation}
where,
\begin{eqnarray}
  a_0 &=&  4 \left(\chi_1^2-2\right) \left(\chi_1^2+1\right) \left(\chi_1^2+4\right), \nonumber\\
  a_1 &=&  -3 \chi_1^2 \left(13 \chi_1^4-4 \chi_1^2+16\right),  \nonumber\\
  a_2 &=& 12 \chi_1^4 \left(\chi_1^2-2\right) ,  \nonumber\\
  a_3 &=&  -4 \chi_1^6.
\end{eqnarray}
With \emph{Mathematica} this equation can be solved exactly. Two roots are imaginary, and only one is real. The threshold $\sigma(\chi_1)$ thus defined is plotted in Figure \ref{fig:sigmaseuil} and its full expression is reported in Appendix \ref{ap}, Eq. \ref{eq:chiStab}. In the strong shock limit $\chi_1 = \infty$, the stability frontier reaches the asymptotic value $\sigma_a$,
\begin{equation}\label{eq:sigma_a}
\sigma_a=1-\frac{3}{2} \sqrt[3]{\sqrt{2}+1}+\frac{3}{2 \sqrt[3]{\sqrt{2}+1}} \sim 0.106.
\end{equation}
The stability threshold attains $\sigma=0$ for $\chi_1 = \sqrt{2}$. A Taylor expansion near $\chi_1=\sqrt{2}$ gives,
\begin{equation}
\sigma = \frac{2^{3/2}}{5} (\chi_1-\sqrt{2} ) + \mathcal{O} (\chi_1-\sqrt{2} )^2.
\end{equation}
The frontier reaches a maximum for $\chi_1=2.42$ and $\sigma_c=0.14$.

\begin{figure}
\begin{center}
 \includegraphics[width=.45\textwidth]{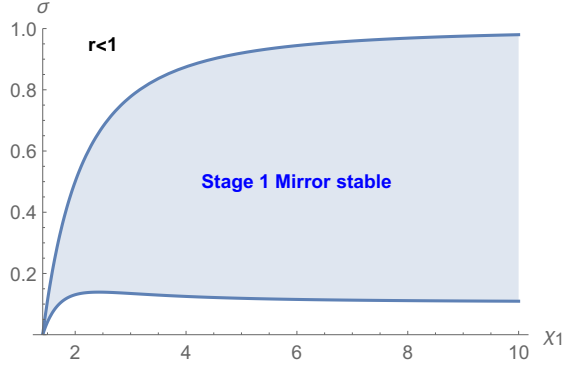}
\end{center}
\caption{Domain of the $(\chi_1,\sigma)$ phase space where Stage 1 defines a shock and is stable. Below the lower frontier, defined by Eq. (\ref{eq:BetaPara2}), the field is too weak to stabilize the anisotropy. Above the upper frontier, defined by Eq. (\ref{eq:NoShock}), the field switches off the shock.}\label{fig:sigmaseuil}
\end{figure}

We also picture on Fig. \ref{fig:sigmaseuil} the limit (\ref{eq:NoShock}) beyond which $r<1$. Stage 1 has stable solutions only in the shaded region.

We therefore find that Stage 1 ($T_{2y}=T_1$) always has $A_2 > 1$ and can be stabilized with a magnetic field, as shown by the shaded region in Figure \ref{fig:sigmaseuil}. If it is mirror unstable, then the downstream plasma will migrate toward the mirror stability threshold. We shall now see that in this case, the conservation equations determine uniquely all of the downstream properties.

\section{Stage 2}\label{sec:stage2}
In case Stage 1 is unstable, it has $A_2>1$ (see Fig. \ref{fig:A2CGL}), so the downstream plasma moves toward the mirror threshold. We therefore impose now,
\begin{equation}\label{eq:StabS2}
A_2=1+\frac{1}{\beta_{2\parallel}}.
\end{equation}
In order to compute the density jump, we start again from,
$$U_2=\frac{1}{2}(k_BT_{2y}+2k_BT_{2x}).$$
Being on the mirror threshold, Eq. (\ref{eq:StabS2}) imposes $T_{2y}=T_{2x}-B_2^2/8\pi n_2$, so that,
\begin{equation}
U_2=\frac{1}{2}\left(k_BT_{2x}-\frac{B_2^2/8\pi}{n_2} +2k_BT_{2x}\right) = \frac{1}{2n_2}\left(3P_{2x}-\frac{B_2^2}{8\pi}\right).
\end{equation}
We then apply the same method than before, replacing $U_2$ in Eq. (\ref{eq:conser4}) by the expression above. We find the following 3rd degree polynomial equation for the density jump $r$,
\begin{eqnarray}\label{eq:rseuil}
P(r)&=&2 \chi_1^2 ~ r^3 +  \left(\frac{10}{\sigma }+\frac{2 \chi_1^2}{\sigma }-4 \chi_1^2\right)~ r^2   \nonumber \\
     && - \left(\frac{10}{\sigma }   +\frac{10 \chi_1^2}{\sigma }+5 \chi_1^2\right)~ r  \nonumber \\
     && + \frac{8 \chi_1^2}{\sigma }=0.
\end{eqnarray}
\begin{figure}
\begin{center}
 \includegraphics[width=.45\textwidth]{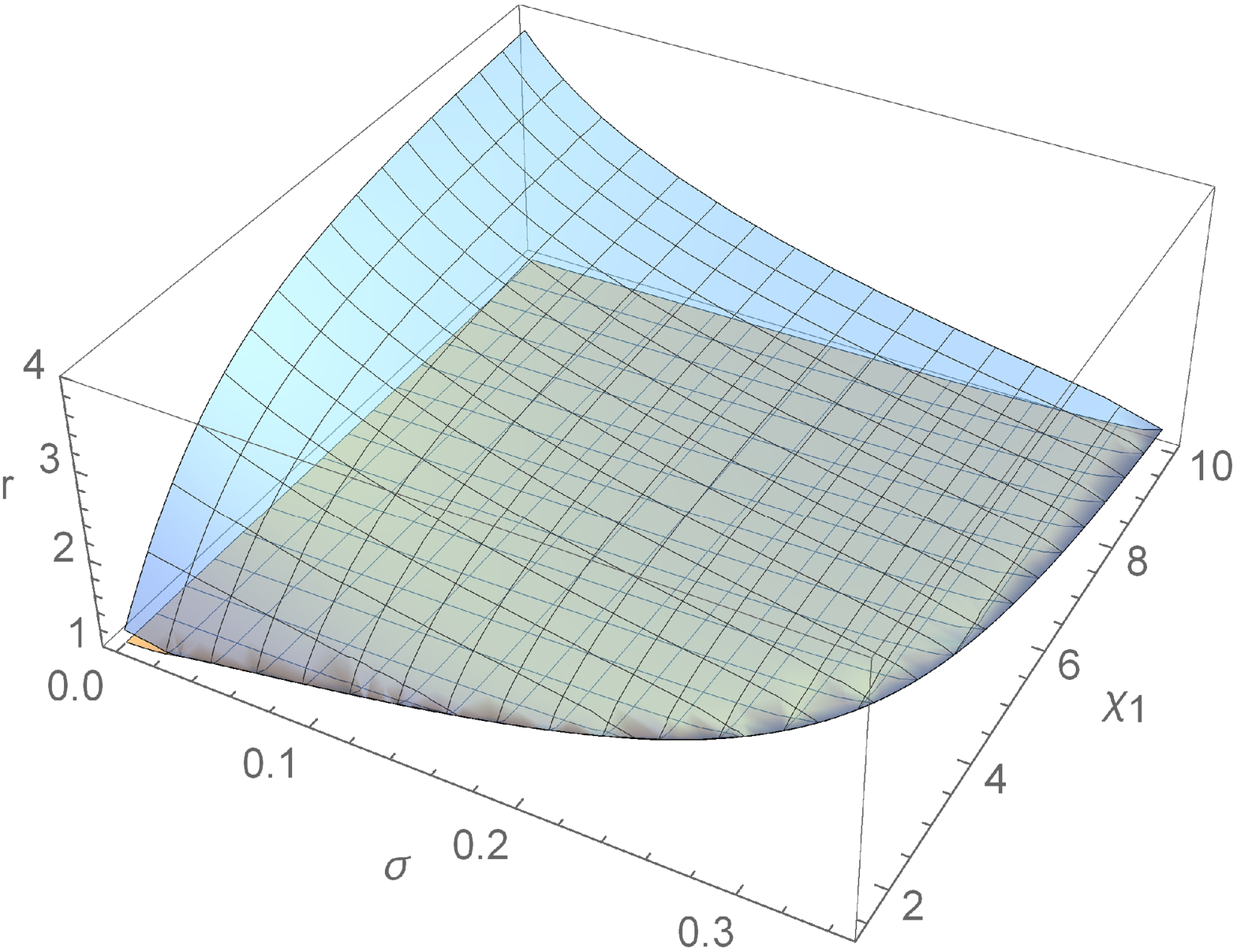}  \includegraphics[width=.45\textwidth]{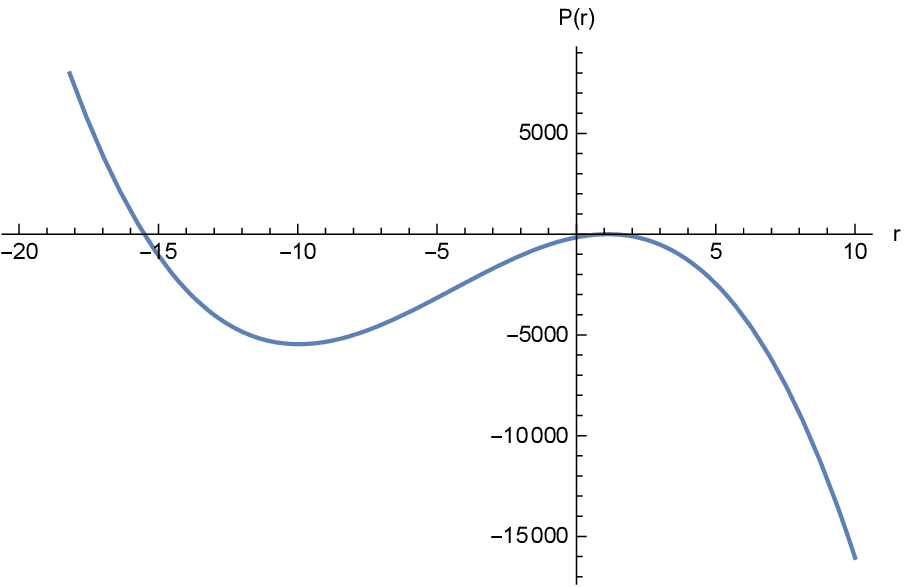}
\end{center}
\caption{\emph{Left}: Plot of the 2 real roots of Eq. (\ref{eq:rseuil}). The physical one is the upper one, which merges with the MHD solution for $\sigma=0$. \emph{Right}: Typical shape of $P(r)$, here with $\sigma=0.2$ and $\chi_1=2$.}\label{fig:seuil}
\end{figure}
We now determine upon which conditions on $(\sigma, \chi_1)$ this equation offers real solutions for the density jump $r$.

As an even degree polynomial, it has always at least one real root. Indeed, we shall see that there are either 1 or 3 real roots.  In case there are 3 roots, 2 of them are $>0$ and one is $<0$. Figure \ref{fig:seuil} plots the 2 positive roots in terms of $(\chi_1,\sigma)$. They join on a frontier studied below, and the physical one is the upper one as it merges with the MHD solution for $\sigma=0$.

In order to make sense of the result, it is useful to further study the polynomial (\ref{eq:rseuil}). Let us symbolically write it as,
$$P(r)=ar^3+br^2+cr+d,$$
and denote its 3 roots $r_1,r_2,r_3$. They fulfill the identities,
\begin{eqnarray}
  \prod_i r_i &=& -d/a=-4/\sigma < 0 \label{eq:root1} \\
  \sum_i r_i  &=& -b/a = -2-\frac{1}{\sigma}\left( 1 + \frac{5}{\chi_1^2}  \right) < 0. \label{eq:root2}
\end{eqnarray}
In addition, one can compute $\partial P/\partial r$ and show it has always 2 purely real roots. Because $a<0$, the shape of $P(r)$ is typically like the one pictured on Fig. \ref{fig:seuil}-Right. As long as we have 3 real roots, 2 of them are positive, and the third has to be negative to fulfill (\ref{eq:root1}). When the value of the right extremum falls below 0, the 2 corresponding real roots become imaginary conjugate and the third one remains negative.

\begin{figure}
\begin{center}
 \includegraphics[width=.45\textwidth]{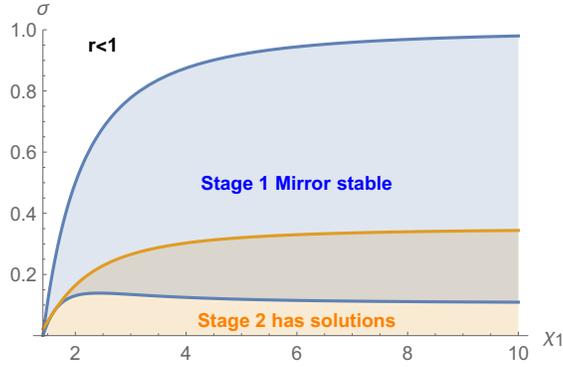}
\end{center}
\caption{Stage 1 is stable and defines a shock between the two blue lines. Stage 2 offers solutions below the orange line.
The orange curve is always above the lower blue one, even at low $\chi_1$, where they are exactly tangent (at least  numerically) to each other for $\chi_1=1.6$. }\label{fig:Stage12}
\end{figure}

Stage 2 offers therefore solutions as long as this right extremum on Fig. \ref{fig:seuil}-Right is larger than 0. Since $\partial P/\partial r$ is of 2nd order, it can be solved exactly, giving the values of $r_\pm$ so that $\partial P/\partial r (r_\pm) =0$. The largest extremum of $P(r)$ is found at,
\begin{eqnarray}
r_+ &=& \frac{(4 \sigma +2) \chi_1^2+10+\sqrt{\Delta}}{6 \sigma  \chi_1^2}, ~~\mathrm{with}  \\ \nonumber
\Delta &=& \left(46 \sigma ^2+76 \sigma +4\right) \chi_1^4+20 (7 \sigma +2) \chi_1^2+100.
\end{eqnarray}
The equation $P(r_+)=0$ then gives the region of the $(\chi_1,\sigma)$ phase space where Stage 2 offers solutions. This region is plotted in Figure \ref{fig:Stage12} together with the stability region of Stage 1. There is a significant overlap between the two domains. Namely, there is a $(\chi_1,\sigma)$ domain where Stage 1 is stable while Stage 2 already offers solutions. In such a case and according to the kinetic history we hypothesized, the downstream should settle in Stage 1 since it first goes through this stage.

\begin{figure}
\begin{center}
 \includegraphics[width=0.45\textwidth]{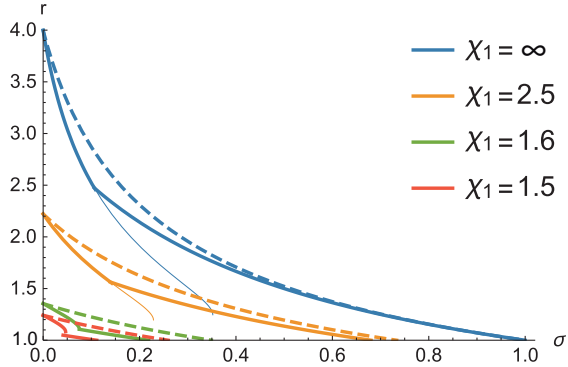}
\end{center}
\caption{Density jump in terms of $\sigma$. At low $\sigma$, Stage 2 offers solutions. At large $\sigma$, the jump is given by Stage 1, which is stable. When both stages offer solutions at once, the physical solution is Stage 1. This is the reason why part of the Stage 2 solutions (low $\sigma$), are pictured in thin lines. The dashed lines picture the ``isotropic'' MHD result (\ref{eq:jumpMHD}).}\label{fig:JumpStage12}
\end{figure}

\section{Putting the 2 stages together}\label{sec:together}
We can now put the 2 stages together and plot the jump in terms of $\sigma$. This is done in Figure \ref{fig:JumpStage12}. It is  interesting to compare with the result (\ref{eq:jumpMHD}) of ``isotropic'' MHD. The dashed lines on Fig. \ref{fig:JumpStage12} picture the ``isotropic'' MHD result (\ref{eq:jumpMHD}).

At low $\sigma$, the field is too weak to stabilize Stage 1 so that the system ends in Stage 2. Then the field becomes strong enough to stabilize Stage 1, even though Stage 2 still offers solutions. In that case, the system settles in Stage 1, since this is the first stage of its kinetic history. As a result, the corresponding part of the jump for Stage 2 are in thin lines on the figure. Then for even larger fields, Stage 2 no longer has solutions while Stage 1 is stable. There, the jump in unambiguously given by Stage 1.

As a consequence of the downstream anisotropy allowed by the field, the density jump is smaller than in isotropic MHD. Yet, unlike the jump reduction in the parallel case which can reach 50\%, the difference here is minor for at least two reasons. To start with, the isotropic MHD jump decreases with the field whereas it is independent from the field in the parallel case. Then, the microphysical explanation highlighted in the parallel case could also play a role. More precisely, the downstream anisotropy in the parallel case is related to the effect of the flow aligned field which guides the particles in the downstream, preventing isotropization \cite{BretJPP2016,BretJPP2017}. In the present perpendicular case, the field rather helps the shock formation instead of hindering it in the parallel case (see conclusion).

\section{Conclusion}
We have investigated the departure from MHD of the density jump of a non relativistic perpendicular shock. Such a departure comes from a pressure anisotropy in the downstream (the upstream is assumed isotropic). Vlasov theory alone cannot pinpoint any definite downstream anisotropy. It only allows for a range of stable anisotropic plasmas instead. In order to derive a theory of the downstream anisotropy in terms of the field, we made an ansatz on the kinetic history of the plasma as it crosses the shock front.

The departure from MHD is less pronounced than in the parallel case, consistently with what is expected from collisionless shock formation theory. Indeed, when two collisionless plasma shells collide, they overlap and the overlying region becomes unstable to competing streaming instabilities \cite{BretPRL2008,BretPoPReview}. The shock starts forming when the turbulence arising from the growth of  instabilities becomes capable of blocking the incoming flow \cite{BretPoP2013,BretPoP2014,RuyerPoP2015,Stockem2015ApJ,RuyerPRL2016}. Yet, a parallel magnetic field will tend to guide the particles in the overlapping region, hindering the density build up. On the contrary, a perpendicular field will help the particles to stay in the overlapping region, contribution to the density build up.

Our model requires $\chi_1^2 > 2$. Since the upstream Mach number $M_1$ verifies $M_1^2=\chi_1^2/\gamma$, choosing $\gamma=5/3$ imposes $M_1 > \sqrt{2/\frac{5}{3}}=1.1$. Indeed, the assumed kinetic history yields no shock solution for $1<M_1<1.1$. As evidenced by Eq. (\ref{eq:NoShockMHD}) and Figure \ref{fig:jumpMHD}-\emph{right}, this stands in contrast with the MHD case where solutions are available from $M_1>1$. This difference is only notable for weak fields as both models share the same shock existence criteria for $\sigma \sim 1$, i.e, $\chi_1 \gg 1$, as can be seen comparing Eqs. (\ref{eq:NoShockMHD}) and (\ref{eq:NoShock}). Our analysis shows that in the weak shock limit, the conservation equations forbid the conservation of the temperature parallel to the field.

Future works contemplate the extension to the relativistic regime, the exploration of oblique field orientations or PIC simulations aiming at assessing the assumed kinetic history of the plasma. To our knowledge, no PIC simulations have been performed in the strong field regime, namely $\sigma \sim 1$. While our scenario should hold in the limit $\sigma = \infty$, PIC simulations will be needed to assess how instabilities in the shock transition, for example, could affect the result for $\sigma  \sim 1$ for the present perpendicular case, and for $\sigma \gg 1$ in the parallel case.

\section{Acknowledgments}
A.B. acknowledges support by grants ENE2016-75703-R from the Spanish
Ministerio de Educaci\'{o}n and SBPLY/17/180501/000264 from the Junta de
Comunidades de Castilla-La Mancha. R.N. acknowledges support from the NSF
via grant AST-1816420. A.B. thanks the Black Hole Initiative (BHI) at Harvard
University for hospitality, and R.N. thanks the BHI for support. The BHI is funded
by a grant from the John Templeton Foundation.

\appendix

\section{Stability threshold for Stage 1}\label{ap}
Stage 1 discussed in Section \ref{sec:stage1} is stable for strong enough a field. It turns mirror unstable for $\sigma$ lower than,
\begin{equation}\label{eq:chiStab}
\sigma(\chi_1) = \frac{\mathcal{A}-\mathcal{B}+2 \left(\chi _1^2-2\right)}{2 \chi _1^2},
\end{equation}
with,
\begin{eqnarray}
  \mathcal{A} &=& \frac{3^{2/3} \chi _1 \left(3 \chi _1^2+4\right)}{\mathcal{C}} \nonumber\\
  \mathcal{B} &=&  3^{1/3}  \chi _1 \mathcal{C} \nonumber\\
  \mathcal{C} &=& \left[9 \chi _1 (\chi_1^2-2) + \sqrt{6} \sqrt{27 \chi _1^6+126 \chi _1^2+32} \right]^{1/3}  \nonumber
\end{eqnarray}
%


\begin{thebibliography}{10}

\bibitem{Zeldovich}
I.A.B. Zel'dovich and Y.P. Raizer.
\newblock {\em Physics of Shock Waves and High-Temperature Hydrodynamic
  Phenomena}.
\newblock Dover Books on Physics. Dover Publications, 2002.

\bibitem{PRLBow1}
S.~D. Bale, F.~S. Mozer, and T.~S. Horbury.
\newblock Density-transition scale at quasiperpendicular collisionless shocks.
\newblock {\em Phys. Rev. Lett.}, 91:265004, Dec 2003.

\bibitem{PRLBow2}
Steven~J. Schwartz, Edmund Henley, Jeremy Mitchell, and Vladimir
  Krasnoselskikh.
\newblock Electron temperature gradient scale at collisionless shocks.
\newblock {\em Phys. Rev. Lett.}, 107:215002, Nov 2011.

\bibitem{Sagdeev66}
R.~Z. {Sagdeev}.
\newblock {Cooperative Phenomena and Shock Waves in Collisionless Plasmas}.
\newblock {\em Reviews of Plasma Physics}, 4:23, 1966.

\bibitem{Stockem2012}
Anne Stockem, Frederico Fi\'{u}za, Ricardo~A Fonseca, and Luis~O Silva.
\newblock The impact of kinetic effects on the properties of relativistic
  electron-positron shocks.
\newblock {\em Plasma Physics and Controlled Fusion}, 54:125004, 2012.

\bibitem{Caprioli2014}
D.~Caprioli and A.~Spitkovsky.
\newblock Simulations of ion acceleration at non-relativistic shocks. i.
  acceleration efficiency.
\newblock {\em Astrophys. J.}, 783:91, 2014.

\bibitem{CGL1956}
G.~F. Chew, M.~L. Goldberger, and F.~E. Low.
\newblock The boltzmann equation and the one-fluid hydromagnetic equations in
  the absence of particle collisions.
\newblock {\em Proceedings of the Royal Society of London A: Mathematical,
  Physical and Engineering Sciences}, 236(1204):112--118, 1956.

\bibitem{BretJPP2017}
Antoine {Bret}, Asaf {Pe'er}, Lorenzo {Sironi}, Aleksander {S\c{a}dowski}, and
  Ramesh {Narayan}.
\newblock {Kinetic inhibition of magnetohydrodynamics shocks in the vicinity of
  a parallel magnetic field}.
\newblock {\em Journal of Plasma Physics}, 83:715830201, 2017.

\bibitem{BretJPP2016}
A.~Bret.
\newblock Particles trajectories in weibel magnetic filaments with a
  flow-aligned magnetic field.
\newblock {\em Journal of Plasma Physics}, 82:905820403, 2016.

\bibitem{Gary1993}
S.P. Gary.
\newblock {\em Theory of Space Plasma Microinstabilities}.
\newblock Cambridge Atmospheric and Space Science Series. Cambridge University
  Press, 1993.

\bibitem{MarucaPRL2011}
B.~A. Maruca, J.~C. Kasper, and S.~D. Bale.
\newblock What are the relative roles of heating and cooling in generating
  solar wind temperature anisotropies?
\newblock {\em Phys. Rev. Lett.}, 107:201101, Nov 2011.

\bibitem{SchlickeiserPRL2011}
R.~Schlickeiser, M.~J. Michno, D.~Ibscher, M.~Lazar, and T.~Skoda.
\newblock Modified temperature-anisotropy instability thresholds in the solar
  wind.
\newblock {\em Phys. Rev. Lett.}, 107:201102, Nov 2011.

\bibitem{Karimabadi95}
H.~Karimabadi, D.~Krauss-Varban, and N.~Omidi.
\newblock Temperature anisotropy effects and the generation of anomalous slow
  shocks.
\newblock {\em Geophysical Research Letters}, 22(20):2689--2692, 1995.

\bibitem{Erkaev2000}
N.~V. {Erkaev}, D.~F. {Vogl}, and H.~K. {Biernat}.
\newblock {Solution for jump conditions at fast shocks in an anisotropic
  magnetized plasma}.
\newblock {\em Journal of Plasma Physics}, 64:561--578, November 2000.

\bibitem{Vogl2001}
D.~F. Vogl, H.~K. Biernat, N.~V. Erkaev, C.~J. Farrugia, and S.~M\"uhlbachler.
\newblock Jump conditions for pressure anisotropy and comparison with the
  earth's bow shock.
\newblock {\em Nonlinear Processes in Geophysics}, 8(3):167--174, 2001.

\bibitem{Gerbig2011}
D.~Gerbig and R.~Schlickeiser.
\newblock Jump conditions for relativistic magnetohydrodynamic shocks in a
  gyrotropic plasma.
\newblock {\em The Astrophysical Journal}, 733(1):32, 2011.

\bibitem{BretJPP2018}
Antoine {Bret} and Ramesh {Narayan}.
\newblock {Density jump as a function of magnetic field strength for parallel
  collisionless shocks in pair plasmas}.
\newblock {\em Journal of Plasma Physics}, 84:905840604, Dec 2018.

\bibitem{Guo2017}
X.~{Guo}, L.~{Sironi}, and R.~{Narayan}.
\newblock {Electron Heating in Low-Mach-number Perpendicular Shocks. I. Heating
  Mechanism}.
\newblock {\em \apj}, 851:134, December 2017.

\bibitem{Guo2018}
X.~{Guo}, L.~{Sironi}, and R.~{Narayan}.
\newblock {Electron Heating in Low Mach Number Perpendicular Shocks. II.
  Dependence on the Pre-shock Conditions}.
\newblock {\em \apj}, 858:95, May 2018.

\bibitem{2019NatAstro}
Marco {Miceli}, Salvatore {Orlando}, David~N. {Burrows}, Kari~A. {Frank},
  Costanza {Argiroffi}, Fabio {Reale}, Giovanni {Peres}, Oleh {Petruk}, and
  Fabrizio {Bocchino}.
\newblock {Collisionless shock heating of heavy ions in SN 1987A}.
\newblock {\em Nature Astronomy}, 3:236--241, Jan 2019.

\bibitem{Gary2009}
S.~P. {Gary} and H.~{Karimabadi}.
\newblock {Fluctuations in electron-positron plasmas: Linear theory and
  implications for turbulence}.
\newblock {\em Physics of Plasmas}, 16(4):042104, April 2009.

\bibitem{Lichnerowicz1976}
A.~{Lichnerowicz}.
\newblock {Shock waves in relativistic magnetohydrodynamics under general
  assumptions}.
\newblock {\em Journal of Mathematical Physics}, 17:2135--2142, 1976.

\bibitem{LandauKinetic}
L.~D. Landau and E.~M. Lifshitz.
\newblock {\em Course of Theoretical Physics, Physical Kinetics}, volume~10.
\newblock Elsevier, Oxford, 1981.

\bibitem{Kulsrud2005}
Russell~M Kulsrud.
\newblock {\em Plasma physics for astrophysics}.
\newblock Princeton Univ. Press, Princeton, NJ, 2005.

\bibitem{FeynmanVol2}
R.P. Feynman, R.B. Leighton, and M.L. Sands.
\newblock {\em The Feynman Lectures on Physics}.
\newblock Number v. 2 in The Feynman Lectures on Physics.
  Pearson/Addison-Wesley, 1963.

\bibitem{Marcowith2016}
Alexandre Marcowith, Antoine Bret, A~Bykov, Mark~Eric Dieckman, Luke Drury,
  Bertrand Lemb\`{e}ge, Martin Lemoine, Giovanni Morlino, Gareth Murphy, Guy
  Pelletier, Ilya Plotnikov, Brian Reville, Mario Riquelme, Lorenzo Sironi, and
  Anne {Stockem Novo}.
\newblock The microphysics of collisionless shock waves.
\newblock {\em Reports on Progress in Physics}, 79:046901, 2016.

\bibitem{fitzpatrick2014plasma}
R.~Fitzpatrick.
\newblock {\em Plasma Physics: An Introduction}.
\newblock Taylor \& Francis, 2014.

\bibitem{BretPRL2008}
A.~Bret, L.~Gremillet, D.~B\'{e}nisti, and E.~Lefebvre.
\newblock Exact relativistic kinetic theory of an electron-beam-plasma system:
  Hierarchy of the competing modes in the system-parameter space.
\newblock {\em Phys. Rev. Lett.}, 100:205008, 2008.

\bibitem{BretPoPReview}
A.~Bret, L.~Gremillet, and M.~E. Dieckmann.
\newblock Multidimensional electron beam-plasma instabilities in the
  relativistic regime.
\newblock {\em Phys. Plasmas}, 17:120501, 2010.

\bibitem{BretPoP2013}
A.~{Bret}, A.~{Stockem}, F.~{Fiuza}, C.~{Ruyer}, L.~{Gremillet}, R.~{Narayan},
  and L.~O. {Silva}.
\newblock {Collisionless shock formation, spontaneous electromagnetic
  fluctuations, and streaming instabilities}.
\newblock {\em Physics of Plasmas}, 20(4):042102, April 2013.

\bibitem{BretPoP2014}
A.~Bret, A.~Stockem, R.~Narayan, and L.~O. Silva.
\newblock Collisionless weibel shocks: Full formation mechanism and timing.
\newblock {\em Physics of Plasmas}, 21(7):072301, 2014.

\bibitem{RuyerPoP2015}
C.~Ruyer, L.~Gremillet, A.~Debayle, and G.~Bonnaud.
\newblock Nonlinear dynamics of the ion weibel-filamentation instability: An
  analytical model for the evolution of the plasma and spectral properties.
\newblock {\em Physics of Plasmas}, 22(3):032102, 2015.

\bibitem{Stockem2015ApJ}
A.~{Stockem Novo}, A.~{Bret}, R.~A. {Fonseca}, and L.~O. {Silva}.
\newblock {Shock Formation in Electron-Ion Plasmas: Mechanism and Timing}.
\newblock {\em Astrophysical Journal Letters}, 803:L29, 2015.

\bibitem{RuyerPRL2016}
C.~Ruyer, L.~Gremillet, G.~Bonnaud, and C.~Riconda.
\newblock Analytical predictions of field and plasma dynamics during nonlinear
  weibel-mediated flow collisions.
\newblock {\em Phys. Rev. Lett.}, 117:065001, Aug 2016.

\end{thebibliography}

\end{document}